\input{aipcheck}

\documentclass[
    ,final            
  ]
  {aipproc}

\usepackage{amsmath,amssymb,amsfonts,latexsym,cancel} 

\layoutstyle{6x9}

\begin{document}

\title{Internal Heating of Old Neutron Stars: Contrasting Different Mechanisms}



\author{Denis Gonz\'alez}{
  address={Departamento de Astronom\'\i a y Astrof\'\i sica, Pontificia Universidad
        Cat\'olica de Chile, Casilla 306, Santiago 22, Chile.}
}

\author{Andreas Reisenegger}{
  address={Departamento de Astronom\'\i a y Astrof\'\i sica, Pontificia Universidad
        Cat\'olica de Chile, Casilla 306, Santiago 22, Chile.}
}

\author{Rodrigo Fern\'andez}{
  address={ Institute for Advanced Study, Princeton, NJ 08540, USA }
}

\begin{abstract}
	The thermal emission detected  from the millisecond pulsar J0437-4715 is not explained
	by standard cooling models of neutron stars without a heating mechanism. We investigated
	three heating mechanisms controlled by the rotational braking of the pulsar: breaking of the solid crust, superfluid
        vortex creep, and non-equilibrium reactions (``rotochemical heating''). We find that the crust cracking mechanism  does not
	produce detectable heating. Given the dependence of the heating mechanisms on spin-down parameters, which leads to different 
	temperatures for different pulsars, we study the thermal evolution for two types of pulsars: young, slowly rotating  ``classical'' 
	pulsars  and old, fast rotating millisecond pulsars (MSPs). We find that the rotochemical heating and vortex creep mechanism 
	can be important both for classical pulsars and MSPs. 
\end{abstract}
\maketitle
\section{INTRODUCTION}

	For all standard cooling models \citep{yak04}, neutron stars (NSs) cool down to surface
        temperatures $T_s < 10^4$ K within less than $10^7$ yr. However, the observation of ultraviolet thermal
        emission from millisecond pulsar J0437-4715 \citep{kar04}, whose spin-down age
        is $\tau_{sd}\sim 7\times 10^9$ yr  \citep{del08}, suggests a surface temperature $\sim 10^5$ K for this pulsar. 
	Thus, a heating mechanism has to be added to the thermal evolution models in order to obtain agreement between theory and 
	observation.

        The goal of this work is  to provide a comparative analysis of the thermal evolution including
        three internal heating mechanisms that are driven by the rotational evolution: crust cracking \citep{baym71,chen92}, 
	vortex creep \citep{alp84a,shi89,lar99}, and rotochemical heating \citep{reis95}.
        The first operates in the nuclear lattice that composes the crust of the star; the second, in the inner crust, where superfluid 
	neutrons coexist with the nuclear lattice; and the last one, mainly in the core of the star.
        Due to the dependence of heating mechanisms on spin-down parameters, we study separately the thermal evolution for two
        types of pulsars: ``classical'' pulsars, characterized by strong magnetic fields and relatively slow rotation, and MSPs
        with low surface magnetic fields and fast rotation.
\section{Internal heating mechanisms}
        Typically, NSs are  born with temperatures  $\sim10^{11}$K. The subsequent cooling is caused by  neutrino emission from the 
	interior and photon emission from the surface of the star. The evolution of the internal temperature is given by the thermal 
	balance equation
        \begin{equation}
        \dot T= \frac{1}{C} ( L_H - L_{\gamma} - L_{\nu}),
        \end{equation}
        where $C$ is the total heat capacity of the star, $L_{\gamma}$ is the photon luminosity, $L_{\nu}$ is the neutrino luminosity,
	and $L_H$ is the power generated by internal heating mechanisms, of which we consider the following three:\\
{\bf Crust Cracking}: The rotational braking causes a gradual change in the shape of the star from an ellipsoidal to a
	spherical form,  which increases the stress in the crust. When it reaches a critical deformation, it breaks,  and part of the 
	accumulated strain energy is released. The energy loss rate by this mechanism, based on the formalism of \citet{baym71}, is 
        $L_{cc}=5bI\theta_c\Omega \dot \Omega$,
        where $b$ is the rigidity parameter of the crust, $I$ is the moment of inertia of the star, $\theta_c$ is the maximum strain 
	angle, and $\Omega$ is the angular velocity of the NS. With $b\sim 10^{-7}$ \citep{cut03} and  $\theta_{c}\sim 10^{-1}$  
	\citep{hor09}, we find $L_{cc}\sim 10^{26} \textrm{erg s}^{-1}$ for typical MSPs, too low to produce detectable 
	heating.\\
{\bf Vortex Creep}: When a superfluid is forced to rotate, this generates vortex lines  whose distribution and 
	dynamics are determined by the rotation rate. In a rotating NS containing a neutron superfluid, the vortices  interact with 
	the nuclei of the inner crust  and are dragged through the nuclear lattice due to the spin-down of the star. The pinning
	and unpinning of the vortex lines with respect to the nuclei of the crystal lattice releases energy that heats the star. 
	The energy  dissipation rate  by this mechanism is given by 
        $L_{vc}= J \dot \Omega$ \citep{alp84a},
        where the parameter $J$ contains all the (highly uncertain) information about the vortex-nuclei interaction.\\
{\bf Rotochemical Heating}: In chemical equilibrium, in a NS composed of neutrons ($n$), protons ($p$), and leptons ($l$: 
	electrons and muons), the chemical potentials $\mu_i$ satisfy $\eta_{npl}\equiv \mu_{n}-\mu_{p}-\mu_{l}=0$. 
	However, if the rotation of the star is slowing down, the centrifugal force is reduced, the central density of the star increases, 
	and the chemical potentials are imbalanced, $\eta_{npl} \neq0$. As the equilibrium composition is altered, the NS will tend to 
	relax to the new chemical equilibrium (via beta and inverse beta decays), releasing energy as neutrinos, antineutrinos, 
	and thermal photons. The evolution of the chemical imbalance is of the form
$\dot \eta_{npl}=-A(\eta_{npl},T)-R_{npl}\Omega\dot\Omega$ \citep{fer05},
\begin{figure}[t]
\includegraphics[width=9cm]{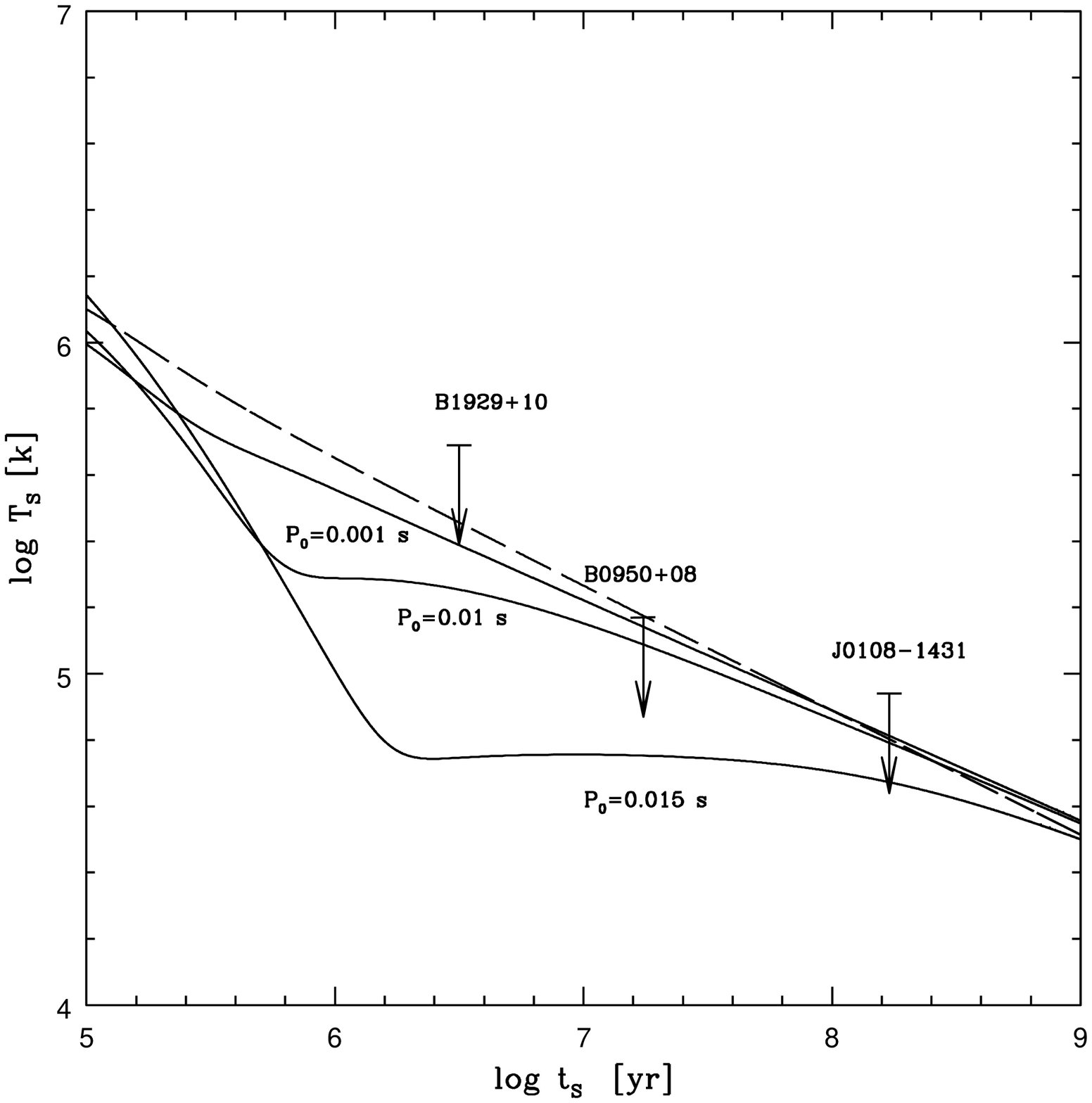}
\includegraphics[width=9cm]{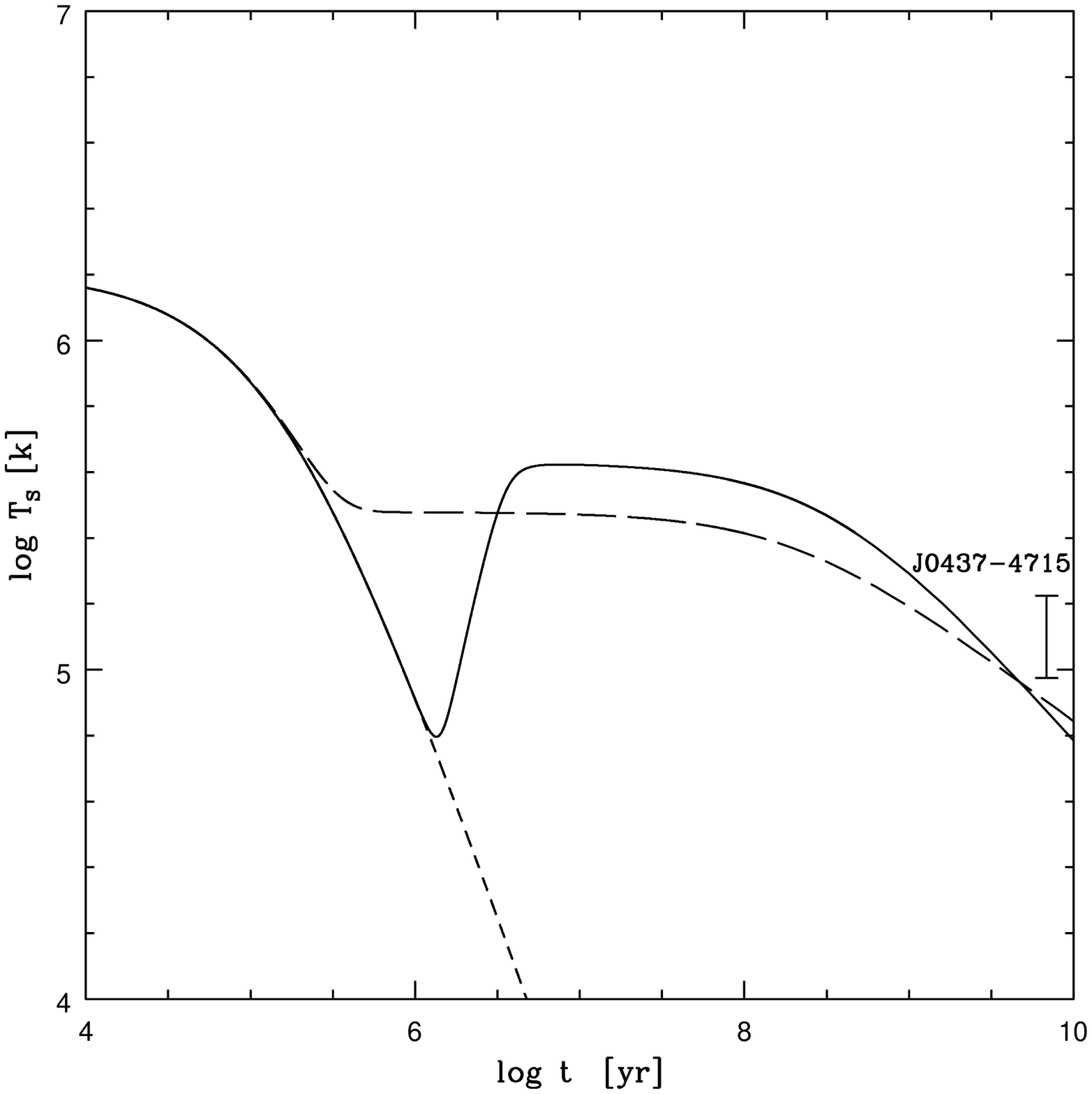}
\caption{ 
{\bf Left panel}: Thermal evolution with rotochemical (solid lines) and vortex-creep (long-dashed line) heating in classical pulsars.
 All curves correspond to stars with  mass $M=1.4M_{\odot}$, magnetic field $B=2.5\times10^{11}\rm{G}$, A18 + $\delta \upsilon$ + UIX* EOS,
and initial temperature $T=10^{11}\rm{K}$. The abscissa correspond to  the spin-down time ($t_s$), and the temperature upper limits 
are observational 
constraints for specific pulsars.
{\bf Right panel}:  MSP  with mass $M=1.76M_{\odot}$, magnetic field $B=2.8\times10^8\rm{G}$, initial
period $P_0=1\rm{ms}$, and A18+$\delta v$+UIX* EOS. The solid and long-dashed curves correspond to rotochemical and vortex creep heating,
respectively. The short-dashed curve shows the thermal evolution  without heating (standard cooling). The error bar shows
the temperature measured for the MSP J0437-4715.}
\end{figure}
	where the function $A$ quantifies the effect of reactions toward restoring chemical equilibrium, and the constant $R_{npl}$ 
	quantifies  the departure from equilibrium due to  the change in the angular velocity $\Omega$ of the star.  \citet{reis95} 
	found that, if the angular velocity $\Omega$ varies slowly over the times required to cool the star and achieve chemical 
	equilibrium, the star reaches a quasi-steady state, with heating and cooling balancing each other. \citet{fer05} calculated 
	the simultaneous solution of  $\dot T = \dot \eta_{npl} = 0$ for a typical range of equations of state and found that, in a 
	NS with a non-superfluid core, the photon luminosity in the quasi-steady state depends only on the period and  period 
	derivative  of the star, $L_{\gamma}^{st} \simeq (10^{30}-10^{31})({\dot{P}_{-20}}/{P_{\rm{ms}}^3})^{8/7}\textrm{erg s}^{-1}$.
        Here, $\dot P_{-20}$ is the period derivative measured in units of $10^{-20}$ and $P_{\rm{ms}}$ is the period in milliseconds.


\section{ Results and Discussion}

	In order to solve the thermal balance equation (1) and generate the evolutionary curves, we 
	use the computational code of \citet{fer05}. This considers a realistic model of NS structure, with
	the core composed of  neutrons, protons,  electrons
	and muons. In order to model the rotational evolution, it is assumed that the rotational braking is due to the 
	 magnetic dipole radiation, without magnetic field decay.

	Thus, using an equation of state (EOS) that allows only modified Urca reactions, A18 + $\delta 
       \upsilon$ + UIX* EOS, which makes the  cooling relatively slow in comparison to EOSs that allow direct Urca reactions,
	Figure 1, left panel, shows the thermal evolution including vortex creep and rotochemical heating
	in  classical pulsars. In the vortex creep mechanism, due to the uncertain value of the pinning energies for the interaction 
	vortex-nuclei in the inner crust,  the parameter $J$ is adjusted to the temperature upper limit of pulsar 
	B0950+08, wich constrains it to 
	$J< 2.8\times 10^{43} \textrm{erg s}$. As the luminosity of this mechanism depends only  on the
	angular velocity derivative $\dot \Omega$, the resulting thermal evolution, for ages $> 10^7 \rm{yr}$,  
	is independent of the initial conditions and the previous 
	thermal history. On other hand, in the case of rotochemical heating, the high magnetic field of classical pulsars, 
	$B > 10^{11}\rm{G}$, 
	causes strong braking of the pulsar, inducing high chemical imbalances in the beginning of the thermal evolution
	that later are slowly reduced  by  chemical reactions. 
	Because of this, in the classical pulsar regime, the thermal evolution with rotochemical heating is very dependent on the 
	uncertain initial rotation period $P_0$ of NSs.

	Similarly, the right panel shows the thermal evolution for MSPs.
	As in the classical pulsars, the vortex creep heating
        is independent of initial conditions or previous thermal history.  
	In this case, 	we assumed that the parameter $J$ is universal. 
	Thus, in the thermal evolution,	
	we used the previous upper limit on $J$ imposed by the pulsar B0950+08, resulting in a ``phenomenological''
	temperature prediction very near that observed in the pulsar J0437-4715.  
	On the other hand, for rotochemical heating, due to the relatively weak magnetic field, $B\sim 10^8 \rm{G}$, the chemical
	imbalances 
	induced by the rotational braking grow slowly, generating chemical reactions at  high ages, $t\gtrsim 10^7 \rm{yr}$ \citep{fer05}.
	Through this, the star arrives at a quasi-steady state in which cooling and heating are balanced, and where the thermal evolution
	is independent of initial conditions and only depends on the  current value of the product $\Omega\dot\Omega$.

	So, we find that the vortex creep and rotochemical heating  can be important both for classical pulsars and MSPs.
	The temperature observed in the MSP J0437-4715 is slightly above the temperature prediction for both mechanisms. 
	However, the temperature predictions of rotochemical heating can be raised if a superfluid
	core is considered in the models \citep{petro}. Finally, better constraints on the temperature of some classical pulsars
	such as B0950+08
	could rule out vortex creep  heating as the  generator of the thermal emission detected
	in the MSP J0437-4715.


\begin{theacknowledgments}
This work was supported by Proyecto Regular FONDECYT 1060644, the FONDAP Center of Astrophysics (15010003), Fellowship from 
ALMA-CONICYT Project (31070001), Proyecto Basal PFB-06/2007, and GEMINI-CONICYT project (32080004).

\end{theacknowledgments}



\bibliographystyle{aipproc}   


\IfFileExists{\jobname.bbl}{}
 {\typeout{}
  \typeout{******************************************}
  \typeout{** Please run "bibtex \jobname" to optain}
  \typeout{** the bibliography and then re-run LaTeX}
  \typeout{** twice to fix the references!}
  \typeout{******************************************}
  \typeout{}
 }

\end{document}